\documentclass[12pt,letterpaper]{article}
\usepackage[utf8]{inputenc}
\usepackage{amsmath}
\usepackage{amsfonts}
\usepackage{amssymb}
\usepackage{graphicx}
\usepackage{slashed}
\usepackage[left=3.00cm, right=3.00cm, top=3.00cm, bottom=3.00cm]{geometry}
\begin{document}
\title{Minkowski-space solutions of the Schwinger-Dyson equation for the fermion propagator with the rainbow-ladder truncation}
\author{Shaoyang Jia\footnote{E-mail: sjia@iastate.edu}, Pieter Maris\\
{\small\textit{Department of Physics and Astronomy, Iowa State University, Ames, Iowa 50011, USA}} \and
Dyana C. Duarte, Tobias Frederico, Wayne de Paula\\
{\small\textit{Instituto Tecnol\'ogico da Aeron\'autica, DCTA, 12.228-900 S\~ao Jos\'e dos Campos, Brazil}} \and 
Emanuel Ydrefors\\
{\small\textit{Institut de Physique Nucl\'eaire, Universit\'e Paris-Sud, }}\\
{\small\textit{IN2P3-CNRS, 91406 Orsay Cedex, France }}\\
{\small\textit{Instituto Tecnol\'ogico da Aeron\'autica, DCTA, 12.228-900 S\~ao Jos\'e dos Campos, Brazil}}}
\maketitle
\begin{abstract}
We solve the Minkowski-space Schwinger-Dyson equation (SDE) for the fermion propagator in quantum electrodynamics (QED) with massive photons. Specifically, we work in the quenched approximation within the rainbow-ladder truncation. Loop-divergences are regularized by the Pauli-Villars regularization. With moderately strong fermion-photon coupling, we find that the analytic structure of the fermion propagator consists of an on-shell pole and branch-cuts located in the timelike region. Such structures are consistent with the direct solution of the fermion propagator as functions of the complex momentum. Our method paves the way towards the calculation of the Minkowski-space Bethe-Salpeter amplitude using dressed fermion propagator.
\end{abstract}

\section{Introduction to the Schwinger-Dyson equation for the fermion propagator in QED with massive photons}
The Schwinger-Dyson equations (SDEs) are coupled integral equations of the Green's functions in a quantum field theory. They serve as recurrences relations of the $n$-point expansions of the generating functional, therefore are inherently nonperturbative. Albeit traditionally solved in the Euclidean space, we expect these integral equations to be well-defined for the Green's functions in the Minkowski momentum space~\cite{Sauli:2004bx,Jia:2017niz} extending to the whole complex momentum plane~\cite{Frederico:2019noo}. In this proceeding, we solve the fermion propagator directly in the Minkowski space from its SDE.

Let us start with the Lagrangian of quantum electrodynamics (QED) with massive photons in the $R_\xi$ gauge~\cite{peskin2018introduction} 
\begin{equation}
\mathcal{L}=\overline{\psi}(i\slashed{D}-m_{\mathrm{B}})\psi +\frac{1}{2}A_\mu\left[g^{\mu\nu}\left(\partial^2+ m_{\mathrm{A}}^2 \right)-\left(1-\frac{1}{\xi} \right)\partial^\mu \partial^\nu  \right]A_{\nu}.\label{eq:lagrangian}
\end{equation}
Here $\psi$ is the fermion field with bare mass $m_{\mathrm{B}}$. $A^\mu$ is the massive gauge boson field with bare mass $m_{\mathrm{A}}$. The covariant derivative is defined as $D^\mu =\partial^\mu -ieA^\mu$ with $e$ being the elementary charge. From Eq.~\eqref{eq:lagrangian}, the bare propagator of the gauge boson is given by
\begin{equation}
D^0_{\mu\nu}(q)=\left[g_{\mu\nu}+(\xi-1)q_\mu q_\nu /(q^2-\xi m_{\mathrm{A}}^2)\right]/(q^2-m_{\mathrm{A}}^2+i\varepsilon).
\end{equation}
In the quenched approximation, the gauge-boson propagator remains bare. The metric tensor is defined as $g^{\mu\nu}=\mathrm{diag}\{1,-1,-1,-1\}$ such that $q^2>0$ corresponds to $q^\mu$ being timelike. 

The fermion propagator $S_{\mathrm{F}}(p)$ is to be solved from its SDE given by $S^{-1}_{\mathrm{F}}(p)={\slashed{p}-m_{\mathrm{B}}-\Sigma(p)}$, with the fermion self-energy $\Sigma(p)$ defined as
\begin{equation}
\Sigma(p) \equiv\slashed{p}\Sigma_{\mathrm{v}}(p^2)+\Sigma_{\mathrm{s}}(p^2) = -ie^2\int d\underline{k}\,\gamma^\nu\,S_{\mathrm{F}}(k)\Gamma^{\mu}(k,p)D_{\mu\nu}(q).\label{eq:def_fermion_self_energy}
\end{equation}
Here we define the integral measure as $\int d\underline{k}={\int d^4k/(2\pi)^4}$. In Eq.~\eqref{eq:def_fermion_self_energy}, we have decomposed $\Sigma(p)$ into its Dirac vector part $\Sigma_{\mathrm{v}}(p^2)$ and Dirac scalar part $\Sigma_{\mathrm{s}}(p^2)$. Meanwhile, $\Gamma^\mu(k,p)$ is the one-particle-irreducible fermion-photon vertex, which in the rainbow-ladder truncation becomes $\gamma^\mu $. 

\section{The spectral representation of the fermion propagator}
Following the procedures in Ref.~\cite{Delbourgo:1977jc}, the K\"{a}ll\'{e}n-Lehmann spectral representation of the scalar propagator is generalized to the fermion propagator: 
\begin{equation}
S_{\mathrm{F}}(p)\equiv S_{\mathrm{v}}(p^2)+S_{\mathrm{s}}(p^2)=\int_{m^2}^{+\infty}ds\dfrac{\rho_{\mathrm{v}}(s)}{p^2-s+i\varepsilon}+\int_{m^2}^{+\infty}ds\dfrac{\rho_{\mathrm{s}}(s)}{p^2-s+i\varepsilon}. \label{eq:SR_S_vs}
\end{equation}
The parameter $m$ is the fermion on-shell mass, which can be solved from ${m\,[1-\Sigma_{\mathrm{v}}(m^2)]}= {m_{\mathrm{B}}+\Sigma_{\mathrm{s}}(m^2)}$. Equation~\eqref{eq:SR_S_vs} is equivalent to $S_{\mathrm{F}}(p)=\int dW\rho(W)/{[\slashed{p}-W+i\,\mathrm{sign}(W)\,\varepsilon]}$, with the support of the integral being ${(-\infty,\,-m]\cup [m,\,+\infty)}$ and $\rho(W)$ defined by $\rho(W)=\mathrm{sign}(W){[W\,\rho_{\mathrm{v}}(W^2)+\rho_{\mathrm{s}}(W^2)]}$. 

Equations~\eqref{eq:SR_S_vs} indicates that both $S_{\mathrm{v}}(p^2)$ and $S_{\mathrm{s}}(p^2)$ are holomorphic functions of the complex momentum except for $p^2\geq m^2$. We further assume that singularities of $S_{\mathrm{F}}(p)$ specified by $\rho(W)$ are given by a mass-shell pole and branch-cuts along the positive-real axis. Such an assumption allows us to write $\rho_j(s)={Z\,\delta(s-m^2)}+{r_j(s)}$ for $j\in\{\mathrm{v},\mathrm{s}\}$. The first term contributes to a simple pole structure of the propagator with residue $Z=\left[1-\Sigma_{\mathrm{v}}(m^2) \right]^{-1}$. After defining $A(p^2)=1-{\Sigma{\mathrm{v}}(p^2)}$ and $B(p^2)=-m_{\mathrm{B}}-{\Sigma_{\mathrm{s}}(p^2)}$, the second term $r_j(s)$ as the branch-cut part of the spectral functions can be calculated from 
\begin{equation}
\begin{cases}
r_{\mathrm{v}}(s)=-\dfrac{1}{\pi}\mathrm{Im}\bigg\{ \dfrac{A(s)}{A^2(s)p^2-B^2(s)} \bigg\}\bigg\vert _{s\geq p^2_{\mathrm{th}}}\\[4mm]
r_{\mathrm{s}}(s)=-\dfrac{1}{\pi}\mathrm{Im}\bigg\{ \dfrac{-B(s)}{A^2(s)p^2-B^2(s)} \bigg\}\bigg\vert _{s\geq p^2_{\mathrm{th}}}
\end{cases}.
\end{equation}
Here $p^2_{\mathrm{th}}$ is the threshold above which the imaginary parts of the fermion self-energy become nonzero. Specifically, we work with massive photons in the Feynman gauge such that there is a finite distance between the pole and the branch points. 

\section{Solution of the SDE in the quenched approximation}
Within the quenched approximation and the Feynman gauge $(\xi=1)$, we calculate the fermion self-energy from its definition in Eq.~\eqref{eq:def_fermion_self_energy}. The momentum dependence of $\Sigma(p)$ is given by 
\begin{equation}
\dfrac{\delta\,\Sigma(p)}{\delta\,\rho(W)}=-ie^2\int d\underline{l}\int dF^2\,\dfrac{-2y\slashed{p}+4W}{[l^2-(xW^2+ym_{\mathrm{A}}^2-xyp^2)+i\varepsilon]^2},\label{eq:Sigma_F_loop_int}
\end{equation}
with $\int dF^2=\int_{0}^{1}dx\int_{0}^{1}dy\,\delta(1-x-y)$. Based on Eq.~\eqref{eq:Sigma_F_loop_int}, the real part of the fermion self-energy is logarithmic divergent, which we remove using the Pauli-Villars regularization. Specifically, the regularized fermion self-energy $\Sigma_{\mathrm{PV}}(p)$ is defined as 
\begin{equation}
\Sigma_{\mathrm{PV}}(p) =\Sigma(p)-(m_{\mathrm{A}}\rightarrow \Lambda),\label{eq:def_PV_reg}
\end{equation}
with $\Lambda>m_{\mathrm{A}}$ being the mass of the regulator. 

Because of Eq.~\eqref{eq:def_PV_reg}, $\Sigma_{\mathrm{PV}}(p)$ vanishes asymptotically. This results in an additional benefit of the Pauli-Villars regularization that the direct implementation of the spectral representation for the self-energy is allowed. Consequently, the imaginary parts of the fermion self-energy uniquely determine $\Sigma_{\mathrm{PV}}(p)$ in the complex momentum plane. Furthermore, we can calculate the imaginary parts of the fermion self-energy in terms of the spectral functions of the fermion propagator using
\begin{align}
-\dfrac{1}{\pi}\mathrm{Im}\big\{ \Sigma_{\mathrm{PV}j}(p^2) \big\} 
&= \dfrac{e^2}{(4\pi)^2}\Bigg\{ \int_{m^2}^{(\sqrt{p^2}-\sqrt{m_{\mathrm{A}}^2})^2}ds\, \mathcal{K}_j(p^2,s,m_{\mathrm{A}})\,\nonumber\\
&\quad\times \sqrt{(p^2-m_{\mathrm{A}}^2+s)^2-4p^2 s} -(m_{\mathrm{A}}\rightarrow \Lambda) \Bigg\}\rho_{j}(s), \label{eq:im_fermion_self_energy}
\end{align}
with $\mathcal{K}_{\mathrm{v}}(p^2,s,m_{\mathrm{A}})=(p^2-m_{\mathrm{A}}^2+s)/p^4$ and $\mathcal{K}_{\mathrm{s}}(p^2,s,m_{\mathrm{A}})=-4/p^2$. 

\begin{figure}
	\begin{center}
		\includegraphics[width=\linewidth]{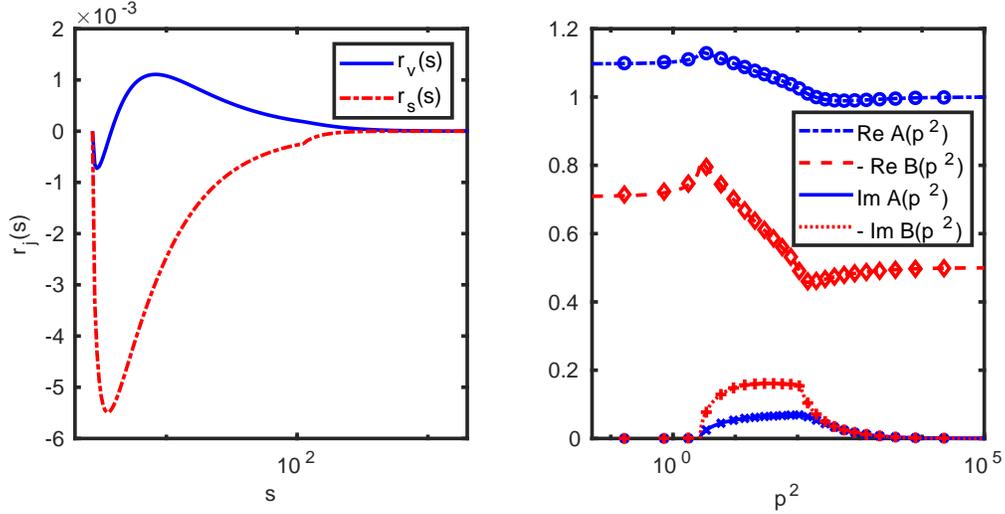}
	\end{center}
	\caption{Solutions of the fermion propagator SDE in massive QED. On the left panel, the blue solid line and the red dash-dot line are the Dirac-vector and the Dirac-scalar parts of the spectral function respectively. On the right panel, lines specified by the legends are the real and the imaginary parts of the inverse propagator. The discretized points correspond to solutions of the same SDE using the technique of un-Wick rotation~\cite{Frederico:2019noo}.}
	\label{fig:spectral_functions}
\end{figure} 
After regularization, the SDE for the fermion propagator becomes a numerically well-defined non-linear integral equation of $\rho_j(s)$. We obtain the numerical solution of the fermion spectral functions through iteration. As an example, we choose the bare mass $m_{\mathrm{B}}=0.5$, the coupling constant $e^2/(4\pi)=0.3$, the vector boson mass $m_{\mathrm{A}}=1$, and the Pauli-Villars mass $\Lambda=10$ to obtain the fermion on-shell mass of $m=0.6502$ with residue $Z=0.9096$. The corresponding branch-cut parts of the spectral functions are illustrated on the left panel of Fig.~\ref{fig:spectral_functions}. With the same input parameters, we also compare the fermion self-energy in the timelike region with results obtained using the technique of un-Wick rotation~\cite{Frederico:2019noo}, as illustrated on the right panel of Fig.~\ref{fig:spectral_functions}. Our results are in agreement within numerical uncertainties in the Euclidean space.
\section{Conclusion}
We have solved the SDE for the fermion propagator in the massive QED in the Landau gauge within the quenched approximation using the Pauli-Villars regularization. With the rainbow-ladder truncation, we find that the structure of the fermion propagator is consistent with a simple pole and bunch-cuts in the timelike region. Our solutions could serve as input conditions for the Minkowski-space Bethe-Salpeter equation of two-fermion bound states. 
\section*{Acknowledgments}
This work was supported by the US Department of Energy under Grants
No. DE-FG02-87ER40371 and No. DE-SC0018223 (SciDAC-4/NUCLEI), and by Funda\c c\~ao de Amparo \`a Pesquisa do Estado de S\~ao Paulo, Brazil (FAPESP) Thematic grants No. 13/26258-4 and No. 17/05660-0, by CAPES, Brazil - Finance Code 001. 
DCD thanks FAPESP grant No.~2017/26111-4, the Simons Foundation under the Multifarious Minds Program grant 557037 and U.S. DOE under Grant No. DE-FG02-00ER41132. 
TF thanks Conselho Nacional de Desenvolvimento Cient\'ifico e Tecnol\'ogico
(Brazil), Project INCT-FNA Proc. No. 464898/2014-5, and the Fullbright Visiting Professor Award. 
WP thanks CNPq under grants 438562/2018-6 and 313236/2018-6 and CAPES under the grant 88881.309870/2018-01. 
EY thanks FAPESP grants No. 2016/25143-7 and No. 2018/21758-2. 
\bibliographystyle{utphys} 
\bibliography{SJ_SDE_bib}
\end{document}